\documentclass[pra,floatfix,twocolumn,showpacs,preprintnumbers,amsmath,amssymb,superscriptaddress]{revtex4}

\usepackage{graphicx}
\usepackage{natbib}
\usepackage{latexsym}
\usepackage{sidecap}
\begin{document}
  
\newcommand{\eich}{\hat{H}}
\newcommand{\ei}{\hat{a}}
\newcommand{\eidag}{\hat{a}^{\dag}}
\newcommand{\bi}{\hat{b}}
\newcommand{\bidag}{\hat{b}^{\dag}}
\newcommand{\ddt}{\frac{d}{dt}}
\newcommand{\Kx}{\hat{K}_x}
\newcommand{\Ky}{\hat{K}_y}
\newcommand{\Kz}{\hat{K}_z}
\newcommand{\Ki}{\hat{K}_i}
\newcommand{\Kj}{\hat{K}_j}
\newcommand{\Kk}{\hat{K}_k}
\newcommand{\Kp}{\hat{K}_+}
\newcommand{\Km}{\hat{K}_-}
\newcommand{\Kpm}{\hat{K}_\pm}
\renewcommand{\exp}[1]{\text{exp}\!\left(\!#1\!\right)}
\newcommand{\bra}[1]{\langle #1|}   
\newcommand{\ket}[1]{|#1\rangle}    
\newcommand{\inp}[2]{\langle #1|#2\rangle}  
\newcommand{\outp}[2]{|#1\rangle \langle #2|}

\title{Quantum Zeno control of coherent dissociation}\author{C. Khripkov}
\affiliation{Department of Chemistry, Ben-Gurion University of
the Negev, P.O.B. 653, Beer-Sheva 84105, Israel}
\author{A. Vardi}
\affiliation{Department of Chemistry, Ben-Gurion University of
the Negev, P.O.B. 653, Beer-Sheva 84105, Israel}

\begin{abstract}
We study the effect of dephasing on the coherent dissociation dynamics of an atom-molecule Bose-Einstein condensate. We show that when phase-noise intensity is strong with respect to the inverse correlation time of the stimulated process,  dissociation is suppressed via a Bose enhanced Quantum Zeno effect.  This is complementary to the quantum zeno control of phase-diffusion in a bimodal condensate by symmetric noise (Phys. Rev. Lett. {\bf 100}, 220403 (2008)) in that the controlled process here is phase-{\it formation} and the required decoherence mechanism for its suppression is purely phase noise. 
\end{abstract}  
\maketitle

Pair production processes are the key to the study of quantum correlations and to the preparation of non-classical states of light and matter.  In optics, eminent correlated-photon experiments \cite{OpticsPairing} have led to the demonstration of Einstein-Podolsky-Rozen (EPR) correlations \cite{EPR}. In matter-wave optics, atom pairs may be produced by dissociation of ultracold diatomic molecules  \cite{Dissociation}, potentially leading to interatomic quantum correlations \cite{CorrelatedAtoms}. In particular the dissociation of a Bose-Einstein Condensate (BEC) of diatomic molecules into boson constituent atoms, is equivalent to optical parametric downconversion in $\chi^{(2)}$ nonlinear Kerr media \cite{BECdiss}, which is the leading mechanism for the generation of correlated photons. The modulational instability implied within this analogy, may enable the generation of pair-correlated or number squeezed matter-wave beams \cite{beams}  and open the way to the realization of Bose-stimulated superchemistry \cite{superchemistry} as well as to novel atom-interferometry techniques \cite{SU11interf}. 

In this work we aim to study the interplay between coherent pair production and decoherence due to the coupling to an external bath. For matter-wave systems such dephasing mechanisms result from inelastic collisions between the Bose condensed atoms and particles (atoms or molecules)  in the surrounding thermal cloud.  It may also be introduced by random variation of the atom-molecule detuning via magnetic field fluctuations in Feshbach-resonance coupled systems \cite{Feshbach} or via laser frequency fluctuations in optically coupled atom-molecule BECs. 

\begin{figure}[t]
\centering
\includegraphics[angle=-90,width=0.45\textwidth]{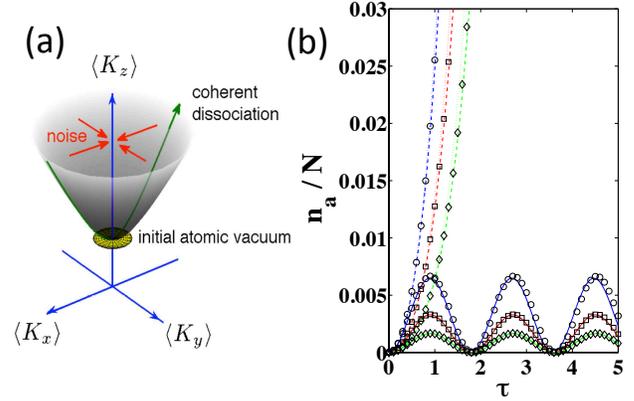}
\caption{(Color online) (a) Noise-control of coherent dissociation. Expectation values for SU(1,1) coherent states, are restricted to the upper-sheet hyperboloid $\langle\Kz^2\rangle - \langle\Kx^2\rangle - \langle\Ky^2\rangle = k^2$. Phase-randomization projects onto the $\Kz$ number-states, thus slowing down the coherent dissociation of the initial atomic vacuum state. (b) Coherent (noise-free) atomic population dynamics, starting from the vacuum $\ket{0,1/4}$ with $\chi=2$ (solid lines), $\chi=0.5$ (dashed lines), and $\chi=0$ (dotted lines). Symbols correspond the linearized analytic expression of Eq. (\ref{spontaneous}). Particle numbers are $N=50$ ($\circ$), 100 ($\square$), and 200 ($\diamond$).}
\label{schematics}
\end{figure}

Rather than study the detrimental effect of weak noise on the dissociation rate, we seek to deliberately {\it use} noise to demonstrate a Bose-stimulated quantum Zeno effect (QZE) \cite{QZE} resulting in the control of the atom-molecule system and the protection the coherent molecular BEC from stimulated dissociation.  As the coherent dissociation is sensitive to the relative phase between the atomic and the molecular modes, the continuous projection onto relative-number states (Fig.~\ref{schematics}(a)) slows down the dissociation. The effect is complimentary to a recent study on the noise-suppression of phase-diffusion in double-well condensates \cite{Khodorkovsky08}. However, whereas site-indiscriminate noise was used in \cite{Khodorkovsky08} to protect against a process where phase-information is {\it lost} in favor of relative-number information, here 'local' dephasing is utilized to arrest phase {\it buildup} by increased number fluctuations. 

We consider the parametric pair production Hamiltonian 
\begin{equation}
\label{Hamilt_ab}
\eich = \frac{\varepsilon}{2}\eidag\ei + \frac{g}{2}(\eidag\eidag\bi + \bidag\ei\ei),
\end{equation}
where $\eidag$ and $\bidag$ denote respectively, creation operators for atomic and molecular modes in the matter wave realization,  $g$ is a measure of the strength of the intermode coupling and $\varepsilon$ is the atomic binding energy in the absence of coupling. The total atom number $N=\hat{n}_a+2\hat{n}_b$, where $\hat{n}_a \equiv \eidag \ei$ and $\hat{n}_b \equiv \bidag \bi$, commutes with the Hamiltonian (\ref{Hamilt_ab}), and is therefore conserved.

The short-time dynamics of molecular dissociation retains SU(1,1) coherence \cite{SU11}. We define the three SU(1,1) generators,
$\Kp \equiv \eidag\eidag/2 $,  $\Km \equiv \ei\ei/2 $, $\Kz=[\Km,\Kp]/2=\eidag\ei/2+1/4$,  
with the complementary operators $\Kx = (\Kp+\Km)/2, \Ky=(\Kp-\Km)/2i$.
The mutual eigenstates of the Casimir operator $\hat{K}^2$ and of $\Kz$ form the number basis set
$\hat{K}^2 \ket{n,k} = k(k-1)\ket{n,k}$,  $\Kz\ket{n,k} = (k+n)\ket{n,k}$,
with $n=0,1,\dots N/2$, and the Bargmann index $k=1/4$ being the minimal value of $\Kz$ when no atoms are present.
The SU(1,1) coherent states $\ket{\theta,\phi}$ are defined in relation to the number basis as
\begin{align}
\label{SU11coherent}
\ket{\theta,\phi} = &\left[1-\mbox{tanh}^2\left(\frac{\theta}{2}\right)\right]^k \sum \limits_{n=0}^{N/2} %
\left[-\mbox{tanh}\left(\frac{\theta}{2}\right)\exp{-i\phi}\right]^n \nonumber \\
&\times\left(\frac{\Gamma(n+2k)}{n!\Gamma(2k)} \right)^{1/2}\ket{n,k}~.
\end{align}
For such states the expectation values $\langle \Ki \rangle\equiv\langle\theta,\phi |\Ki|\theta,\phi\rangle$ are restricted to the upper-sheet of the hyperboloid $\langle\Kz\rangle^2 - \langle\Kx\rangle^2 - \langle\Ky\rangle^2 = k^2$, shown in Fig~\ref{schematics}(a).

Expressing the Hamiltonian in the new operators, we get the simple form
\begin{equation}
\label{Hamilt_su}
\eich = \varepsilon (\Kz - \frac{1}{4}) + g(\Kp\bi + \bidag\Km).
\end{equation}
Starting from the atomic vacuum state $\left |n=0, k=1/4\right\rangle$ corresponding to an all-molecules BEC, it is therefore clear that the state of the system will remain an SU(1,1) coherent state as long as the molecular mode is macroscopically populated and can be replaced by  a classical $c$-number, so that the Hamiltonian (\ref{Hamilt_su}) only generates Lorentzian boosts and rotations. 

Linearization about the atomic vacuum, assuming an undepleted molecular pump $\bi\rightarrow\sqrt{N/2}$ (justified for short time dynamics where the depletion of the pump mode is negligible), results in the atom number evolution \cite{BECdiss},
\begin{equation}
\label{spontaneous}
\frac{\hat{n}_a}{N} = \frac{1}{N(1-\chi^2)}\mbox{sinh}^2\left(\sqrt{1-\chi^2} \tau\right)~,
\end{equation}
where $\chi=\varepsilon/(g\sqrt{2N})$ is the characteristic coupling parameter and time has been rescaled as $\tau=tg\sqrt{N/2}$. Thus, as shown in Fig.~\ref{schematics}(b), for weak atom-molecule coupling ($\chi>1$) the coherent SU(1,1) dynamics correspond to small oscillations of the atomic population, whereas the near-resonance strong coupling ($\chi<1$) evolution corresponds to the Bose-stimulated parametric amplification of emitted atom pairs \cite{BECdiss}.

Regardless of the value of $\chi$, the short-time dynamics of Eq.~(\ref{spontaneous}) corresponds to an initially {\it quadratic} decay with a vanishing rate as $\tau\rightarrow 0$. The timescale on which the buildup of atomic population can be considered quadratic, is the correlation time $\tau_c=(1-\chi^2)^{-1/2}$ corresponding to $t_c=\tau_c/(g\sqrt{N/2})$ of order 1-10 ms for current BEC  setups. This long correlation time enables the realization of a QZE \cite{QZE} which may slow down coherent molecular dissociation. Since the stimulated dissociation process amounts to the formation of a well-defined phase between the atomic and molecular modes at the expense of increased number variance, the desired noise to suppress  it should involve phase fluctuations which will project the state of the system into well-defined relative-number states.

Such phase-noise may be implemented  via stochastic variation of the atom-molecule detuning $\varepsilon$ or via collisions with thermal particles. Controlling the modulation rate or the density of thermal atoms thus provides a handle for controlling dissociation via the noise parameters. In the former case, we stochastically modulate the atom-molecule detuning (e.g. by introducing a fluctuating magnetic field in a Feshbach resonance setup), thus modifying the system Hamiltonian as 
\begin{equation}
\label{Hamilt_S}
\eich_S=\eich+\zeta(t)(\Kz - \frac{1}{4}).
\end{equation}
The stochastic term  $\zeta(t)$ is characterized by its zero mean $\langle{\zeta(t)}\rangle_t=0$ and the characteristic decay time $(t_c)_{noise}$ of the noise correlation function $\langle\zeta(t)\zeta(t')\rangle$. When $(t_c)_{noise}\ll t_c$, we can assume a delta-correlated white noise $\langle\zeta(t)\zeta(t')\rangle\approx\Gamma\delta(t-t')$. The standard iteration of the Liouville equation then results in the Markovian kinetic master equation,
\begin{equation}
\label{master}
\ddt \hat{\rho} = i[\hat{\rho},H] -\Gamma[\hat{\mathcal{L}},[\hat{\mathcal{L}},\hat{\rho}]],
\end{equation}
\noindent where $\hat{\rho}$ is the $N$-particle density matrix of the system, $\Gamma$ is the noise intensity, and the Lindblad noise term $\mathcal{L}$ is the population difference,
\begin{equation}
\label{lindblad}
\hat{\mathcal{L}} = 2\bidag\bi - \eidag\ei = N - 4\Kz + 1.
\end{equation} 

Taking the initial state to be $|0,1/4\rangle$, the undepleted classical-pump approximation amounts to replacing $\bi$ and $\bidag$ with the fixed molecular population $\sqrt{N/2}$. Using Eq.~(\ref{master}) in conjunction with the identities ${\rm Tr}(A[B,C])={\rm Tr}([A,B]C)$, ${\rm Tr}([A,B])=0$, and the SU(1,1) commutation relations, we obtain dynamical equations for the expectation values $K_i\equiv{\rm Tr}\{ {\hat\rho} \Ki\}$:
\begin{eqnarray}
\frac{d}{d\tau} K_x&=&-2\chi K_y - 16\gamma K_x,\nonumber\\
\frac{d}{d\tau} K_y&=&2\chi K_x+2 K_z -16\gamma K_y,\nonumber\\
\label{masterK}
\frac{d}{d\tau} K_z&=&2 K_y,
\end{eqnarray}
where $\gamma\equiv\Gamma/(g\sqrt{N/2})$. Similarly, for the correlation functions $\Delta_{ij}\equiv{\rm Tr}\{ {\hat\rho} (\Ki \Kj+\Kj \Ki)\}-2K_i K_j$ we find,
\begin{eqnarray}
\frac{d}{d\tau} \Delta_{xx}&=&-4\chi\Delta_{xy}-32\gamma\left(\Delta_{xx}-\Delta_{yy}-2K_y^2\right),\nonumber\\
\frac{d}{d\tau} \Delta_{yy}&=&4\chi\Delta_{xy}+4\Delta_{yz}-32\gamma\left(\Delta_{yy}-\Delta_{xx}-2K_x^2\right),\nonumber\\
\frac{d}{d\tau} \Delta_{zz}&=&4\Delta_{yz},\nonumber\\
\frac{d}{d\tau} \Delta_{xy}&=&2\chi\left(\Delta_{xx}-\Delta_{yy}\right)+2\Delta_{xz}-64\gamma\left(\Delta_{xy}+K_x K_y\right),\nonumber\\
\frac{d}{d\tau} \Delta_{xz}&=&-2\chi\Delta_{yz}+2\Delta_{xy}-16\gamma\Delta_{xz},\nonumber\\
\label{BBR}
\frac{d}{d\tau} \Delta_{yz}&=&2\chi\Delta_{xz}+2\left(\Delta_{yy}+\Delta_{zz}\right)-16\gamma\Delta_{yz}~.
\end{eqnarray}

We note that Eqs. (\ref{masterK}) and Eqs. (\ref{BBR}) are valid as long as the molecular field can be assumed to be undepleted ($n_a/N< 0.1$) and do not make any further large-number assumption on the atomic operators $\Ki$. They thus apply also for the very early stages of dissociation when the atomic population comes purely from quantum spontaneous emission. Unlike standard mean-field equations which have the atomic vacuum as an unstable stationary point,  they accurately depict the amplification of $K_z$ and the parametric SU(1,1) phase-squeezing, when $\Gamma=0$.

\begin{figure}[t]
\centering
\includegraphics[angle=0,width=0.48\textwidth]{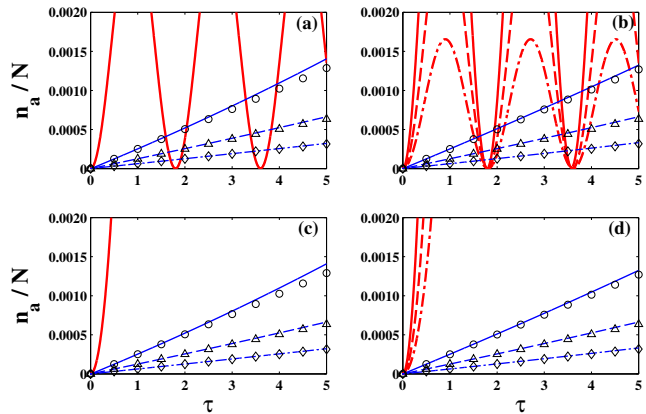}
\caption{(Color online) Atomic population dynamics starting from the atomic vacuum in the presence of noise. Panels (a),(b)  are for subcritical $\chi=2$ whereas panels (c),(d) are for supercritical $\chi=0.5$. Dependence  on noise-intensity is demonstrated in panels (a),(c) where $N=100$ and $\gamma$ is 0 (bold solid line), 5 (solid, $\circ$), 10 (dashed, $\triangle$), and 20 (dash-dotted, $\diamond$). Dependence on particle number is illustrated in panels (b),(d) where $\gamma=0$ dynamics (bold lines) is compared with noise-suppressed dynamics with $\gamma=10$ (normal lines and symbols) for $N=50$ (solid, $\circ$), 100 (dashed, $\triangle$), and 200 (dash-dotted, $\diamond$). Symbols correspond to the analytic QZE expression (\ref{full_na}).}
\label{noisy}
\end{figure}

\begin{figure}[t]
\centering
\includegraphics[angle=0,width=0.4\textwidth]{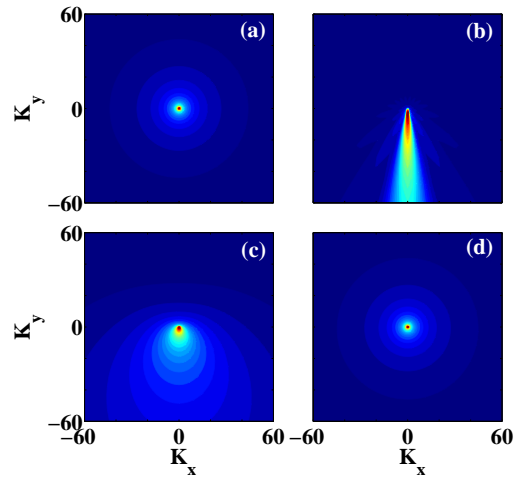}
\caption{(Color online) The SU(1,1) Husimi distribution function $Q(\theta,\phi)$, projected onto the $K_x,K_y$ plane.  The initial state at $\tau=0$ (a) is a coherent vacuum, with equal $\sqrt{N/2}$ variances in $\Kx$ and $\Ky$. In the absence of noise ($\gamma=0$), a definite atom-molecule phase has been acquired at $\tau=3$ (b). By contrast, when noise is introduced  with $\gamma=0.1$ (c) and $\gamma=5$ (d), squeezing is arrested and the initial coherent vacuum is protected. In all plots $\chi=0$ and $N=30$.}
\label{husimi}
\end{figure}

If the frequent measurement condition $\Gamma \gg t_c^{-1}$ is satisfied, $K_x$ and $K_y$ can be adiabatically eliminated in Eqs. (\ref{masterK}), in accordance with the QZE picture where coherence is not allowed to build up due to the introduction of noise. Substituting the initial values corresponding to the all-molecular state ${\bf K}=[0,0,1/4]$, we find that the atomic population is then given by,
\begin{equation}
\label{full_na}
\frac{\hat{n}_a}{N} = \frac{2}{N}\left[ \exp{\frac{\gamma\tau}{\chi^2/4+16\gamma^2}}-1 \right]~,
\end{equation}
At the limit of on-resonance strong interaction, $\chi\rightarrow  0$, this expression reduces to the familiar QZE behavior:
\begin{equation}
\label{qze_na}
\frac{\hat{n}_a}{N} = \frac{2}{N}\left[ \exp{\frac{g^2 N t}{32\Gamma}}-1 \right]~,
\end{equation}
corresponding to the slowing down of stimulated dissociation as the noise intensity $\Gamma$ is increased. Interestingly, since the effective coupling strength $g\sqrt{N}$ is also $N$ dependent, the obtained QZE is Bose enhanced \cite{Khodorkovsky08} as discussed below.

In Fig.~\ref{noisy} we show the early population dynamics obtained from direct numerical simulations of the dissociation process according to Eq.~(\ref{master}). We show representative plots for $\chi$ values above unity (panels a,b) and below it (panels c,d), corresponding to stable and unstable coherent dynamics, respectively. Compare first  the noiseless evolution (solid bold lines) to the QZE-controlled evolution in the presence of noise of varying intensity $\Gamma$ (normal lines), while keeping the particle number $N$ fixed, as shown in Fig.~\ref{noisy}a and Fig.~\ref{noisy}c. Symbols depict the QZE prediction (\ref{full_na}), showing excellent agreement with the exact numerical results. As expected, both stable oscillations and unstable hyperbolic amplification are converted in the presence of noise, to similar $\Gamma$-controlled exponential buildup of atomic population, because the QZE dynamics is composed from repeated $t<t_c$ pieces which have a similar quadratic behavior in both cases.

The dependence of the QZE on particle number is illustrated in Fig.~\ref{noisy}b and Fig.~\ref{noisy}d, where we plot the early dissociation dynamics for different values of $N$, fixing $\chi$  and $\gamma$. The transition from a hyperbolic dissociation rate (\ref{spontaneous}) which depends on $N$ only through its onset time, to the QZE-suppressed dissociation (\ref{qze_na}) which depends linearly on $N$, introduces a factor of $N/\log{N}$ between the coherent and noise-controlled processes. Thus the QZE is strongly Bose-amplified.

\begin{figure}[t]
\centering
\includegraphics[angle=0,width=0.48\textwidth]{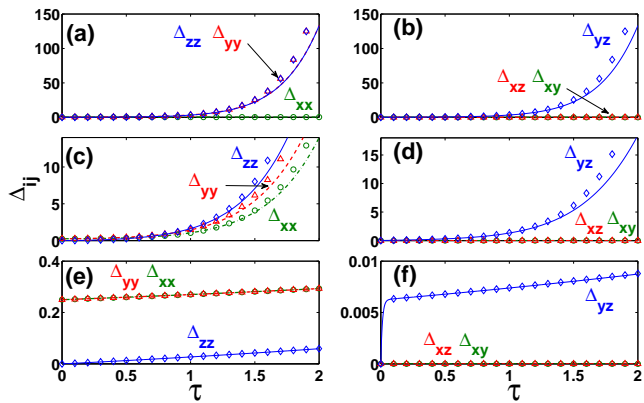}
\caption{(Color online) Dynamics of the correlation functions $\Delta_{ij}$ with $N=100$ and $\chi=0$ starting from $|0,1/4\rangle$, at various  noise intensities: (a,b) $\gamma=0$, (c,d) $\gamma=0.1$, and (e,f) $\gamma=5$. Lines correspond to direct integration of the master equation (\ref{master}), whereas symbols denote the undepleted-pump dynamics of Eqs.~(\ref{BBR}) with initial conditions ${\bf K}=(0,0,1/4)~,~\Delta_{xx}=\Delta_{yy}=1/4~,~\Delta_{zz}=\Delta_{xy}=\Delta_{xz}=\Delta_{yz}=0$, corresponding to the atomic vacuum.}
\label{correlations}
\end{figure}

The effect of dephasing on the coherent dissociation dynamics, is shown in  Fig.~\ref{husimi}, using SU(1,1) Husimi quasi-probability distribution plots. The Husimi distribution is defined in terms of the SU(1,1) coherent states (\ref{SU11coherent}), as $Q(\theta,\phi)\equiv\bra{\theta,\phi}{\hat\rho}(t)\ket{\theta,\phi}$   where ${\hat\rho}(t)$ is the state of  the system at time $t$.  The initial coherent vacuum state at $t=0$ (Fig.~\ref{husimi}(a)) has equal variances in $\Kx$ and $\Ky$. Without noise, it evolves to an SU(1,1) squeezed coherent state (Fig.~\ref{husimi}(b)). We note parenthetically, that unlike SU(2) or Glauber coherent states, SU(1,1) coherence and squeezing are not contradictory \cite{SU11}.  Consequently a definite phase is formed between the atomic and molecular modes. In the presence of noise (Fig.~\ref{husimi}(c),(d)) the squeezing is lowered significantly, with nearly complete protection of the initial molecular state available for sufficiently intense noise.  This picture correlates well with the dynamics of fluctuations presented in Fig.~\ref{correlations}. Without noise (Fig.~\ref{correlations}(a),(b)) we obtain the standard parametric amplification of the number fluctuations $\Delta_{zz}$, accompanied by an equal hyperbolic growth of $K_y$ and $\Delta_{yy}$ at fixed $K_x$ and $\Delta_{xx}$, and leading to decreased phase uncertainty $(\Delta\varphi)^2\propto \Delta_{xx}/(K_x^2+K_y^2)$ \cite{SU11} as illustrated in Fig.~\ref{husimi}(b). In the presence of phase-noise the amplification of $\Delta_{yy}$ is suppressed and $\Delta_{xx}$ begins to grow (Fig.~\ref{husimi}(c) and Fig.~\ref{correlations}(c),(d)) until at the QZE limit (Fig.~\ref{husimi}(d) and Fig.~\ref{correlations}(e),(f)) both transverse fluctuations $\Delta_{xx}$, $\Delta_{yy}$ grow symmetrically at a rate which decreases with $\Gamma$, suppressing the amplification of number fluctuations and the formation of a relative atom-molecule phase. We note that the undepleted-pump correlation equations (\ref{BBR}) agree well with the exact numerical simulations.

To conclude, it is interesting to contrast our results here with previous work on the QZE suppression of phase-diffusion in a bimodal BEC \cite{Khodorkovsky08} with underlying SU(2) spherical algebra. In Ref.~\cite{Khodorkovsky08} we sought to slow down the {\it dephasing} between two BEC modes by tailoring noise which affects both site-modes symmetrically and projects the state of the system onto the initial coherent odd {\it superposition} of the two modes. By contrast, here we aim to slow down the process of {\it phase-formation}, using a much more tenable form of noise, which appears naturally  via collisions or magnetic field fluctuations, and projects the state of the system onto the initial {\it number} state. As anticipated, we have observed that the coherent dissociation of a dimeric molecular BEC consisting of bosonic atoms can be prevented by such 'local' noise.  As in the SU(2) case,  the dissociation becomes less likely as the number of particles increases as $N/\log{N}$. 

This work was supported  by the Israel Science Foundation (Grants 582/07, 346/11) and by grant no. 2008141 from the United States-Israel Binational Science Foundation (BSF).

\end{document}